\documentstyle[stwol]{article}

\input{psfig}

\def\be{\begin{equation}}
\def\ee{\end{equation}}
\def\bea{\begin{eqnarray}}
\def\eea{\end{eqnarray}}

\def\l#1{{\lambda}_{#1}}
\def\lp#1{{\lambda^{\prime}_{#1}}}
\def\lpp#1{{\lambda^{\prime\prime}_{#1}}}

\def\C#1{{C}_{#1}}
\def\Cp#1{{C^{\prime}_{#1}}}
\def\Cpp#1{{C^{\prime\prime}_{#1}}}

\def\be{\begin{equation}}
\def\ee{\end{equation}}
\def\br{\begin{eqnarray}}
\def\er{\end{eqnarray}}

\input psfig

\begin{document}

\title{$R$-PARITY VIOLATION AND FLAVOUR VIOLATION}

\author{ B. DE CARLOS }

\address{
School of Mathematical and Physical Sciences, 
University of Sussex, Falmer, Brighton BN1 9QH, UK}

\author{ P.L. WHITE }

\address{Theoretical Physics, University of Oxford,
1 Keble Road, Oxford OX1 3NP, UK}

\twocolumn[\maketitle\abstracts{
The MSSM with $R$-Parity violation allows flavour changing neutral
currents through two mechanisms, directly through one loop diagrams,
and indirectly through the generation of flavour violation in the
sparticle sector. We discuss the use of these mechanisms for
constraining $R$-Parity violation, and show that the indirect mechanism
parametrised in terms of the running of the soft masses from the
unification or Planck scale typically dominates previously calculated
effects. We discuss neutrino mass generation, $\mu\to e\gamma$, $b\to
s\gamma$, and $K^0-\bar K^0$ mixing as examples.
}]

\section{Introduction}
One of the most promising candidates for physics beyond the so-called 
Standard Model (SM) is that of supersymmetry (SUSY). In this paper we
shall be concerned with the implications of a particular possible
feature of SUSY, namely that of $R$-parity violation
(RPV).\cite{rpv,barbm,suzuki} $R$-parity is a $Z\!\!\! Z_2$ symmetry of
both the SM and its minimal SUSY extension, the MSSM, under which all
of the SM particles have charge 0, while all their SUSY partners have
charge 1. Its implications include the stability of the lightest
supersymmetric particle (LSP), and hence the typical SUSY collider
signatures of missing $E_T$ and the existence of a source of dark
matter. Its violation changes both the implied cosmology and the
expected collider signatures, allowing such effects as LSPs decaying
inside the detector and leptoquarks. In addition to these, further
constraints on RPV can be derived by considering experimental limits on
rare decays.\cite{bgh,bpw,bgnn}

$R$-parity is violated by the superpotential and soft potential terms
\br
W^{RPV}&=&
   \frac{1}{2}\l{ijk}L_iL_je_k + \lp{ijk}L_iQ_jd_k \cr
  && \quad + \frac{1}{2}\lpp{ijk}u_id_jd_k + \mu_iL_iH_2 \cr
V^{RPV}_{\rm soft}&=&
       \frac{1}{2}\C{ijk}L_iL_je_k + \Cp{ijk}L_iQ_jd_k \cr
   && \quad + \frac{1}{2}\Cpp{ijk}u_id_jd_k
             + D_iL_iH_2 \cr
   && \qquad + m^2_{L_ih_1}L_iH_1^* + h.c.
\er
From the point of view of deriving constraints on the $R$-parity
violating couplings in the model, the most extensively studied
couplings are the dimensionless couplings $\l{}$, $\lp{}$, and
$\lpp{}$, which directly generate many effects which can be
experimentally limited. The extra soft terms by definition mostly
couple only heavy SUSY particles and hence are relevant mostly because
of their impact on the RGEs, although they can have significant effects
on the neutrino-neutralino and Higgs-sneutrino
sectors.\cite{suzuki,rv,hemp,rpar}

In SUSY models, flavour changing effects may be caused by the existence
of off-diagonal terms in the sfermion mass matrices in the basis in
which the fermion masses are diagonal. Such flavour-violating soft
masses can be generated either from the high energy theory such as a
GUT directly, or else through the RGEs by couplings which violate
flavour symmetries, such as Yukawa couplings mediated by the CKM
matrix or here RPV couplings.

This paper is a short summary of work contained in two previous
papers,\cite{rpar,qfv} in which we presented the renormalisation group
equations (RGEs) for the couplings of the full $R$-parity violating
sector of the model, and investigated the implications of typical
scenarios at the GUT scale for the generation of neutrino masses and
other flavour violation processes.

\section{Effects of the RGEs}
We begin with a brief discussion of the dimensionless couplings, whose
RGEs have recently been presented in a number of
papers.\cite{rpar,dimless} Here we present simple analytic solutions to
the RGEs in the limit where the Yukawa couplings are much smaller than
the gauge couplings \cite{rpar} leading to 
\begin{eqnarray}
\lambda(M_Z) & = & 1.5\ \lambda(M_{GUT}) \nonumber \\
\lambda^{\prime}(M_Z) & = & 3.4 - 3.7\ 
\lambda^{\prime}(M_{GUT}) \label{LEHE} \\
\lambda^{\prime\prime}(M_Z) & = & 4.0 - 4.7\ 
\lambda^{\prime\prime}(M_{GUT}) \nonumber
\end{eqnarray}
where the ranges are caused by the error on $\alpha_3(M_Z)$. In
addition to running themselves, the RPV couplings can alter the mass
spectrum, giving quite a tight constraint on the $\lp{}$ couplings from
the requirement that the sneutrino mass should be above its
experimental limit,\cite{qfv} and can generate patterns of soft
masses which violate flavour and lepton number symmetries.

\section{Sneutrino VEVs}
Sneutrino VEVs are an important signature of $R$-parity violation 
which arise because of the existence of $\mu_i$, $D_i$, and
$m^2_{L_iH_1}$ terms which explicitly cause the effective potential to
contain terms linear in the sneutrino field, either from explicit
\cite{suzuki,hemp,lee} or spontaneous $R$-parity violation. They can
also be caused by one loop effects involving dimensionless $R$-parity
violating couplings,\cite{bgmt,enqvist} and by generation from the
RGEs \cite{rpar} as discussed below. Once sneutrinos have acquired
VEVs, neutrinos and neutralinos mix, so that we may derive bounds on
$R$-parity violating terms by imposing experimental limits on neutrino
masses, since in general we find that with $l_i$ as the VEV of $L_i$, 
a neutrino mass is generated is of order $(g_1^2+g_2^2)l_i^2/2M$, where
$M$ is some typical neutralino mass.

If we assume universal soft masses, then at the GUT scale we  have only
$R$-parity violation dimensionless and trilinear terms. The dangerous
terms for generating sneutrino VEVs are then generated by the following
terms mixing $L_i$ and $H_1$.
\begin{equation}
\begin{array}{cccl}
\l{i33}h_\tau  & \hbox{or} & \lp{i33}h_b  &
    \hbox{generating $\mu_i$, $D_i$, $m^2_{H_1L_i}$} \\
\C{i33}h_\tau  & \hbox{or} & \Cp{i33}h_b  & 
    \hbox{generating $D_i$} \\
\C{i33}\eta_\tau & \hbox{or} & \Cp{i33}\eta_b  &
    \hbox{generating $m^2_{H_1L_i}$}
\end{array}
\end{equation}
The effects are largest when $\tan\beta$ is large, but the dependence
is rather complicated. The sneutrino VEV will in general be
proportional to the $R$-parity violating coupling, and hence the
neutrino mass to the coupling squared.

We have performed a GUT scale analysis, setting universal parameters at
the unification scale, together with some choice of GUT scale
$R$-parity violating Yukawas, then running masses and couplings to low
energy to give output. Unfortunately, the behaviour is a sufficiently
complicated function of the many different parameters that it is not
really possible to derive useful bounds on the couplings, but it is
nonetheless possible to get an idea of the order of magnitude of the
neutrino mass which we expect. For example we find that $\l{133}$ and
$\lp{133}$ of order $10^{-3}$ and $10^{-4}$ are still large enough to
be inconsistent with present experimental limits over much of parameter
space.\cite{rpar}

\section{Rare and Forbidden Processes}
\subsection{$\mu\to e\gamma$}
One of the most tightly bounded experimental constraints on flavour
changing neutral currents is through the rare decay $\mu\to e\gamma$,
forbidden in the SM. In SUSY models, a non-zero rate can be generated
through non-diagonal slepton mass matrices and also through the direct
effects of $R$-parity violating couplings.\cite{suzuki,lee,bgmt}
However, as noted above, $R$-parity violation induces flavour violation
through soft terms, and so here we shall consider the two effects
together. We shall set only two $R$-parity violating dimensionless
couplings non-zero at $M_{GUT}$ and see what effects they generate. We
will be mainly interested in comparing the relatively simple ``direct''
contributions from diagrams which have $R$-parity violating couplings
at the vertices, with the ``indirect'' contributions where the flavour
violation is driven by off-diagonal mass insertions $\Delta m^2$
generated through the RGEs.

\begin{figure*}[htb]
\center
\psfig{figure=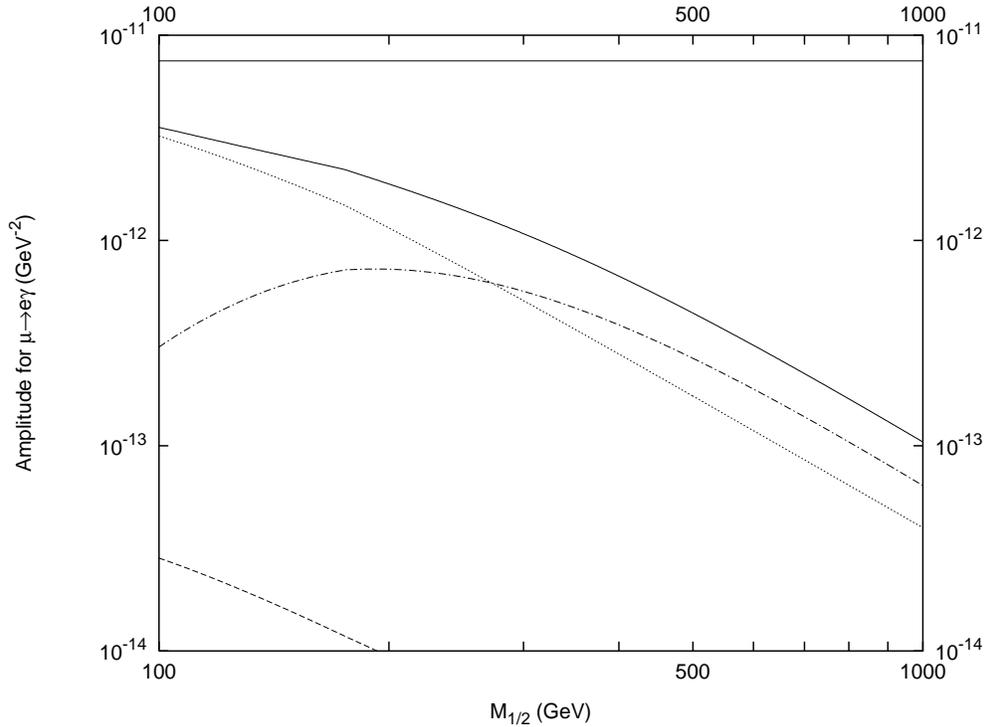,height=10cm}
\caption{
Absolute values of amplitudes for $\mu\to e\gamma$ from direct $R$-parity
violation diagrams (dashed lines), neutralino mediated diagrams
(dot-dashed lines), and chargino mediated diagrams (dotted lines)
plotted against $M_{1/2}$. We also show the total amplitude (solid
line) and the experimental bound on the amplitude (horizontal solid
line). Parameters are $m_t=175$GeV, $\alpha_3(M_Z)=0.12$,
$\tan\beta=10$, $m_0=100$GeV, $A_0=0$, $\mu_4>0$, and
$\lp{111}(M_{GUT})=\lp{211}(M_{GUT})=0.001$.
}
\label{oldfig9}
\end{figure*}

As an example we consider the impact of
$\lp{111}(M_{GUT})=\lp{211}(M_{GUT})=0.001$. We show the resulting
contributions to the amplitude as a function of $M_{1/2}$ in
Figure~\ref{oldfig9}. Here we have set $\tan\beta=10$, $m_0=100$GeV,
$A_0=0$, $\mu_4>0$. What is remarkable about this figure is that it is
clear that in this case the direct contributions are completely
negligible relative to those from the chargino and neutralino mediated
diagrams.

%

We conclude this section by summarising our results. Firstly we find
rather different behaviour for the three different scenarios of
non-zero $\lp{}$, $LH$ $\l{}$ (where the flavour violation occurs in the
left handed slepton sector through $\l{1ij}\l{2ij}$) and $RH$ $\l{}$
(where it occurs in the right handed slepton sector through
$\l{ij1}\l{ij2}$). For the $\lp{}$ case, the chargino and neutralino
mediated diagrams with flavour violation through soft mass insertions
dominate completely the direct contributions, giving very much tighter
constraints, particularly for large $\tan\beta$. For the $LH$ effects
due to $\l{}$ couplings we find again that the chargino contribution
dominates, but not overwhelmingly, and there can be large
cancellations. For the $RH$ case there are no chargino contributions,
and the neutralino and direct effects are usually of comparable size
and opposite sign. However, since there are so many possible
cancellations between terms, it is essentially impossible to derive
concrete bounds. The strongest reasonable statement is that, for the
values we have considered for pairs of couplings at $M_{GUT}$ of
$\lambda\lambda\simeq 10^{-4}$ and $\lp{}\lp{}\simeq 10^{-6}$ we expect
contributions of order the experimental limit for a very light
spectrum, with the branching ratio scaling as $\lambda^4$ or $\lp{}^4$
respectively.

\subsection{$b\to s\gamma$}
Another process which has been studied in the context of constraining
flavour violation in SUSY theories is that of $b\to s\gamma$. Here we
find that the indirect effects again often dominate the direct ones.
However, the bounds on couplings derived here are quite weak, since
$b\to s\gamma$ is much harder to constrain as the large SM contribution
complicates matters, and indeed we find that the bounds on $\lp{}$ are
weaker than those derived from requiring the sneutrino mass to be above
its experimental limit, while those on $\lpp{}$ are only of order 0.2
for the relevant product with a very light spectrum.\cite{qfv}

\subsection{$K^0-\bar K^0$ Mixing}
The final process which we shall consider is that of $K^0-\bar K^0$
mixing, the direct contributions to which have been extensively
studied.\cite{barbm,crs} Here we find a complete contrast to the
situation for the other processes, in that even when
the large tree level contribution is neglected the indirect
contributions are always smaller than the direct ones.\cite{qfv}

\section{Conclusion}
The full RGEs for the MSSM with $R$-parity violation with the inclusion
of all soft terms as well as all dimensionless couplings lead to some
interesting physics. The most important effects of
including $R$-parity violating couplings at the unification scale are
those associated with flavour violation, both through ``direct'' terms
where these couplings appear at the vertices of the diagrams, and
``indirect'' terms where they generate off-diagonal soft masses through
the RGEs which then generate effects through one loop diagrams.

The inclusion of $R$-parity violation in our superpotential through
dimensionless terms allows the generation of lepton-Higgs mixing which
leads to sneutrino VEVs and hence neutrino masses. We have shown that
the indirect generation of sneutrino VEVs through the running of the
RGEs for the soft terms often leads to larger effects than those
derived directly from one loop diagrams. Typically we find that values
of $\l{i33}$ and $\lp{i33}$ of order $10^{-2}$ and $10^{-3}$
respectively at the GUT scale give masses to the corresponding neutrino
of order hundreds to thousands of eV, although the exact value is quite
dependent on the unification scale parameters, and these form the
tightest constraints on these couplings which have been derived.

Similarly, we have studied the process $\mu\to e\gamma$, which we have
shown to be very strongly affected by chargino and neutralino mediated
diagrams. These typically dominate the direct contributions which had
already been calculated, often by several orders of magnitude for the
case of the $\lp{}$ couplings, but there are strong cancellations so
that it is not possible to give precise bounds on the couplings from
such processes. However, unless we invoke arbitrary cancellations, the
typical size of such indirect effects on FCNC are likely to be the
dominant constraint on the building of a model with non-zero $R$-parity
violating couplings. In comparison, bounds derived from $b\to
s\gamma$ are extremely weak except where the spectrum is already
experimentally ruled out, while for $K^0-\bar K^0$ mixing the direct
contributions dominate.

Our main conclusion from these calculations is that $R$-parity
violation can generate large flavour violating effects through the
running of the dimensionful RGEs, and that these effects are often much
larger than those which are generated directly by the couplings
themselves, so that merely studying diagrams with $R$-Parity violating
vertices can be very misleading.

\end{document}